\documentstyle[twoside,fleqn,marco]{article}

\newcommand{\AmS}{{\protect\the\textfont2
  A\kern-.1667em\lower.5ex\hbox{M}\kern-.125emS}}

\newcommand{\beq}{\begin{equation}}
\newcommand{\eeq}{\end{equation}}
\def\beqa{\begin{eqnarray}}
\def\eeqa{\end{eqnarray}}
\def\p{\partial}

\def\lap{\lower.5ex\hbox{$\; \buildrel < \over \sim \;$}}
\def\gap{\lower.5ex\hbox{$\; \buildrel > \over \sim \;$}}

\def\vp{\varphi}

\input{psfig.sty}

\title{Eternal inflation and the present universe}

\author{Alexander Vilenkin \address{Tufts Institute of Cosmology,\\
Physics Department, Tufts University,Medford, MA 02155, USA}
}

\begin{document}

\begin{abstract}
Eternally inflating universes can contain thermalized regions with
different values of the cosmological parameters.  In particular, the
spectra of density fluctuations should be different, because of the
different realizations of quantum fluctuations of the inflaton field.
I discuss a general method for calculating probability distributions
for such variable parameters and analyze the density fluctuation
spectrum as a specific application.
\end{abstract}

\maketitle

\section{Introduction}

In this paper I am going to discuss the structure of the universe on
super-large scales, so large that we are never going to observe them.
I shall argue, however, that this analysis may help us understand some
features of the universe within the observable range.  This is
based on the work done with my student Vitaly Vanchurin at Tufts 
and with Serge Winitzki at Cambridge University.

Let me begin with a brief introduction to eternal inflation.  
As we know, inflation is a nearly exponential expansion of the
universe,
\beq
a(t)\approx e^{Ht},
\label{exp}
\eeq
which is driven by the potential energy of a scalar field $\vp$,
called the inflaton.  $a(t)$ in Eq.(\ref{exp}) is the scale factor and
the expansion rate $H$ is determined by the inflaton potential
$V(\vp)$.  Inflation ends when $\phi$
starts oscillating about the minimum of the potential.  Its energy is
then dumped into relativistic particles and is quickly thermalized.

A remarkable feature of inflation is that generically it never ends
completely.  At any time, there are parts of the universe that are
still inflating \cite{AV83,Linde86}.  
The reason is that the evolution of $\vp$ is
influenced by quantum fluctuations.  This appies in particular to the
range of $\vp$ near the maximum of $V(\vp)$, where the potential is
very flat.   
As a result, thermalization does not occur everywhere at the same time.   
We can introduce a decay constant $\Gamma$ such that $\tau=1/\Gamma$
is the characteristic time it takes $\vp$ to get from the maximum to the
minimum of the potential.  Then the total inflating volume in the
universe is proportional to
\beq
{\cal V}_{inf}\propto e^{-\Gamma t}e^{3Ht}.
\eeq
The first factor on the right-hand side describes the exponential
decay of the inflating volume due to thermalization, while the second
factor describes the exponential expansion of the regions which still
continue to inflate.  For flat potentials required for successful
inflation, we typically have $\Gamma\ll 3H$, so that ${\cal V}_{inf}$
grows exponentially with time.  The thermalized volume grows at the
rate  $d{\cal V}_{therm}/dt = \Gamma d{\cal V}_{inf}/dt$, and thus
${\cal V}_{therm}$ also grows exponentially.

Different thermalized regions in such eternally inflating universe may
have very different properties.  Here are some examples.

The potential $V(\vp)$ may have several minima corresponding 
to vacua with different physical properties. For example, the values
of some constants of Nature (e.g., the electron mass or the
cosmological constant) or cosmological parameters (such as the
amplitude of density fluctuations, the baryon to entropy ratio, etc.) 
could be different in the corresponding
thermalized regions.  A more interesting possibility is that the
``constants'' are related to some slowly-varying fields and take
values in a continuous range.  For example, the inflaton could be a
complex field, $\vp=|\vp|\exp(i\chi)$, with a potential having the
shape of a ``deformed Mexican hat'' (that is, with some $\chi$-dependence).
Then different paths that $\vp$ can take from the top of
the potential to the bottom will result in different
magnitudes of density fluctuations $\delta\rho/\rho$.  The amplitude
of the fluctuations will therefore be different in different parts of
the universe.  Another example is a field $\chi$ (unrelated to the
inflaton) with a
self-interaction potential $U(\chi)$.  If $U(\chi)$ is a very slowly
varying function of $\chi$, then it can act as an effective
cosmological constant.  Quantum fluctuations will randomize $\chi$
during inflation, and observers in different parts of the universe
will measure different values of $U(\chi)$.  

Perhaps the most important example is the spectrum of cosmological
density fluctuations.  The density fluctuation $\delta\rho/\rho(l)$ is
determined by the quantum fluctuation $\delta\vp(l)$ of the inflaton
field $\vp$ at the time when the corresponding comoving scale $l$
crossed the horizon.  Different realizations of quantum fluctuations 
$\delta\vp(l)$ result in different density fluctuations spectra in
widely separated parts of the universe.  This uncertainty is present
in {\it all} models of inflation.

In all these examples, we have parameters $\chi$ which we cannot
possibly predict with certainty.  All we can hope to do is to
determine the probability distribution ${\cal P}(\chi)$.

An eternally inflating universe is inhabited by a huge number of
civilizations that will measure different values of $\chi$.  We can
define the probability ${\cal P}(\chi)d\chi$ as being proportional to
the number of observers who will measure $\chi$ in the interval
$d\chi$ \cite{AV95}.  Now, observers are where galaxies are, and thus 
${\cal P}(\chi)d\chi$ is proportional to the number of galaxies in
regions where $\chi$ takes values in the interval $d\chi$.  
We can then write
\beq
{\cal P}(\chi)\propto F(\chi)\nu(\chi),
\label{PF}
\eeq
where $F(\chi)d\chi$ is the fraction of volume in thermalized regions
with $\chi$ in the interval $d\chi$, and $\nu(\chi)$ is the number of
galaxies per unit volume (as a function of $\chi$).  
The calculation of $\nu(\chi)$ is a standard astrophysical problem,
and here I shall focus on the volume factor $F(\chi)$.

In this discussion I am trying to avoid the word ``anthropic'',
because it makes some people very upset, but what I want to emphasize
is that the approach I have just outlined is as quantitative and
predictive as it can possibly be.  Once ${\cal P}(\chi)$ is
calculated, we can predict, for example, that $\chi$ should have a
value in a certain range with 95\% confidence.

The first attempts to implement this approach encountered an
unexpected difficulty.
It can be traced down to the fact that eternal inflation never
ends, and the number of galaxies in an eternally inflating universe is
infinite at $t\to\infty$.  In order to calculate the volume fraction
$F(\chi)$, one therefore has to compare infinities, which is an
inherently ambiguous procedure.  One can introduce a time cutoff and
include only galaxies that formed prior to some time $t_c$, with the
limit $t_c\to\infty$ at the end.  One
finds, however, that the resulting probability distributions are
extremely sensitive to the choice of the time coordinate $t$
\cite{LLM94,WV96}.  
Linde, Linde and Mezhlumian
\cite{LLM95} attempted to determine the most probable spectrum of
density fluctuations using the proper time along the worldlines of
comoving observers, which they regarded as the most natural choice of
the time coordinate.  They found a probability distribution suggesting
that a typical observer could find herself at a deep minimum of the
density field.  On the other hand, if one uses the expansion factor
along the worldlines as the time coordinate, one recovers the
standard result \cite{Mukhanov}.  Coordinates in general relativity
are arbitrary labels, and such gauge-dependence of the results is, of
course, an embarrassment.

The rest of the paper is organized as follows.  After reviewing the
physics of eternal inflation in Section 2, I discuss the spacetime
structure of an eternally inflating universe in Section 3.  This
will help us understand the origin of the gauge-dependence problem.
The proposed resolution of the problem is discussed in Section 4.  As
a specific application, the spectrum of density fluctuations measured
by a typical observer  is analyzed in section 5.  The conclusions are
briefly summarized in Section 6.

\section{Eternal inflation}

The metric of an inflating universe has a locally Robertson-Walker
form,
\beq
ds^2=dt^2-a^2(t)d{\bf x}^2,
\eeq
with the expansion rate given by
\beq
{\dot a}/a\approx H(\vp)=[8\pi V(\vp)/3]^{1/2}.
\eeq
The potential $V(\vp)$ is assumed to be a slowly varying function of
$\vp$.  As a result, $H$ is a slowly varying function of the
coordinates, and we have an approximately de Sitter space with a
horizon distance $H^{-1}$.  The classical slow-roll evolution equation
for $\vp$ is
\beq
{\dot \vp}_{cl}\approx -H'(\vp)/4\pi.
\label{phieq}
\eeq

Quantum fluctuations of $\vp$ can be represented as a random walk with
random steps taken  
independently in separate horizon-size regions, with one step per
Hubble time $H^{-1}$.
The rms magnitude of the steps is 
\beq
\delta\vp_{rms}=(H/2\pi).
\eeq
We do not have a completely satisfactory derivation of this stochastic
picture in the general case.  Its main justification is that it
reproduces the results of quantum field theory in de Sitter space for
a free scalar field of mass $m\ll H$ (that is, the two-point function
obtained by averaging a classical stochastic field coincides with the
quantum two point function).  For flat inflaton potentials, the
dynamics of $\vp$ should be close to that of a free field, so one
expects the stochastic picture to apply with a good accuracy.

Let us define the distribution 
$F(\vp,t)d\vp$ as the volume occupied by $\vp$ in the
interval $d\vp$ at time $t$.  It satisfies the Fokker-Planck equation
\cite{AV83,Starobinsky,GLM,Sasaki,Bond,LLM94}
\beq
\p_t F+\p_\vp J=3H^\alpha F,
\label{FP}
\eeq
where 
\beq
J=-{1\over{8\pi^2}}\p_\vp(H^{\alpha+2}F)-{1\over{4\pi}}H^{\alpha-1}H'F.
\label{J}
\eeq
The first term of the flux $J$ describes quantum ``diffusion'' of the
field $\vp$, while the second term corresponds to the classical
``drift'' described by Eq.(\ref{phieq}).  The parameter $\alpha$ in
Eqs.(\ref{FP}),(\ref{J}) represents the freedom of time
parametrization, with the time variable $t$ related to the proper time
$\tau$ according to $dt=H^{1-\alpha}d\tau$.  Hence, $\alpha=1$
corresponds to the proper time, $t=\tau$, and $\alpha=0$ to the scale
factor time, $t=\ln a$.  

A great deal of research has been done on the properties of the
Fokker-Planck equation (\ref{FP}) and on its solutions.  To summarize
the conclusions, there are some good news and some bad news.  The good
news is that the asymptotic form of the solutions of (\ref{FP}) is
\beq
F(\phi,t)\to F(\phi)e^{\gamma t} ~~~~~~~(t\to\infty).
\eeq
The overall factor $e^{\gamma t}$ drops out in the normalized
distribution, and thus one gets a stationary asymptotic distribution
for $\vp$.  The bad news is that $F(\vp)$ has a strong dependence on
$\alpha$, so that the results are very sensitive to the choice of the
time coordinate \cite{LLM94}. This is a very disturbing conclusion, and
one could have thought that there is something wrong with the
Fokker-Planck equation (\ref{FP}).  We shall see, however, that there
is a good physical reason for the gauge-dependence of $F(\vp)$, so the
equation is not to blame.

\section{Spacetime structure of an eternally inflating universe}

To analyze the spacetime structure of an eternally inflating universe,
one can perform numerical simulations using the stochastic
representation of quantum fluctuations \cite{Aryal,LLM94}.  
Here, I present the results
of the simulations performed by V. Vanchurin,
S. Winitzki and myself \cite{VVW}.  We consider a comoving region
which has a horizon size $l=H^{-1}$ and a homogeneous inflaton field
$\vp=0$ at the initial moment $t=0$.  We evolve $\vp$ in time
increments $\delta t$ according to the rule
\beq
\delta\vp({\bf x})={\dot\vp}_{cl}({\bf x})\delta t+\delta\vp_q({\bf
x}),
\eeq
where ${\dot\vp}_{cl}$ is from Eq.(\ref{phieq}) and $\delta\vp_q({\bf
x})$ is a random Gaussian field of zero mean,
\beq
\langle \delta\vp_q({\bf x})\rangle=0,
\eeq
and with a correlation function
\beq
\langle \delta\vp_q({\bf x})\delta\vp_q({\bf x'})\rangle
=\left({H\over{2\pi}}\right)^2(H\delta t)C(r).
\eeq
Here, $r$ is the physical distance between the points ${\bf x}$ and
${\bf x'}$, $C(0)=1$, and $C(r)$ rapidly drops to zero at $r \gg
H^{-1}$.  The correct asymptotic form is $C(r)\propto r^{-4}$
\cite{WV99}; in the
simulations we set $C(r)=0$ for $r>2H^{-1}$.  The scale factor $a({\bf
x},t)$ is evolved according to
\beq
\delta a({\bf x})=H({\bf x})a({\bf x})\delta t.
\eeq

The spatial distribution of inflating and thermalized regions in a
$(2+1)$-dimensional simulation is shown in Fig.1.  We used a
double-well potential for the inflaton,
\beq
V(\vp)=V_0\cos^2(\kappa\vp),
\label{Vphi}
\eeq
where $\kappa=\pi/2\eta$, and we only consider the range $-\eta<\vp<\eta$.
Inflating regions are white, and the two types of thermalized regions
corresponding to the minima at $\vp=\pm\eta$ are shown with different
shades of grey.  As time goes on, larger and larger fraction of the
comoving volume gets thermalized, so the inflating regions shrink in
comoving coordinates.  
The inflating regions can be thought of as inflating domain
walls separating the two types of vacua.  Since the domain walls
cannot disappear for topological reasons, it is clear that inflation
must be eternal in this model \cite{LV}.

\begin{figure}[t]
\psfig{file=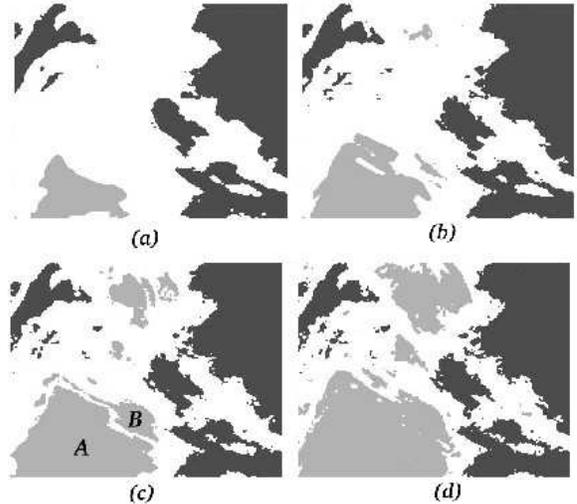,width=3in,angle=0}
\caption{A two-dimensional simulation for the double-well
model at four consecutive moments of time \cite{VVW}.  
We evolved a comoving region of initial size $l=H^{-1}$ with the
initial value of $\phi=0$.  Inflating regions are shown white, while
thermalized regions with $\phi=+\eta$ and $\phi=-\eta$ are shown with
different shades of grey.  Thermalized regions of the same type can
join in the course of the simulation.  For example, regions labeled
$A$ and $B$ in snapshot (c) have merged into a single region in
snapshot (d).  However, regions of different types cannot merge: they
are separated by inflating domain walls.}
\label{fig.1}
\end{figure}

The spacetime structure of the universe in these simulations is
illustrated in Fig.2.  Now, the vertical axis is time and the
horizontal axis is one of the spatial directions.  The boundaries
between inflating and thermalized regions, which play the role of the
big bang for the corresponding thermalized regions, are infinite
spacelike surfaces.  In the figure, these boundaries become nearly
vertical at late times, so that they appear to be timelike.  The
reason is that the horizontal axis in Fig.2 is the comoving distance,
with the expansion of the universe factored out.  The physical
distance is obtained by multiplying by the expansion factor $a(t)$,
which grows exponentially as we go up along the time axis.  If we used
the physical distance in the figure, the thermalization boundaries
would ``open up'' and become nearly horizontal (but then it would be
difficult to fit more than one thermalized region in the figure).

\begin{figure}[t]
\psfig{file=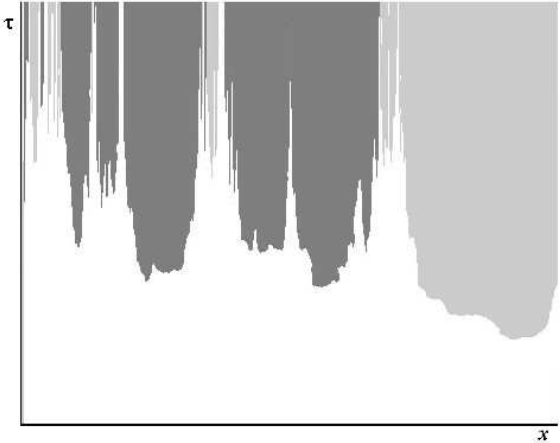,width=3in,angle=0}
\caption{Spacetime structure in a one-dimensional simulation
for the double-well model.  It can be thought of as a spacetime slice
through the $(2+1)$-dimensional simulation illustrated in Fig.1.  The
shading code is the same as in Fig.1.}
\label{fig.2}
\end{figure}

Thermalization is followed by a hot radiation era and then by a
matter-dominated era during which luminous galaxies are formed and
civilizations flourish.  All stars eventualy die, and thermalized
regions become dark, cold and probably not suitable for life.  Hence,
observers are to be found within a layer of finite (temporal)
width along the thermalization boundaries in Fig.2.

It is now easy to see why the probability distributions obtained using
a cutoff at $t=t_c$ are so sensitive to the choice of the time
variable $t$.  Any spacelike surface ${\cal S}$ can be an equal-time
surface $t=t_c$ with an appropriate choice of $t$.  Depending on one's
choice, the surface ${\cal S}$ may cross many thermalized regions of
different types (e.g., for $t=\tau$), may cross only regions of one
type, or may cross no thermalized regions at all (say, for $t=\vp$
with $\vp$ in the deterministic slow-roll range).  These possibilities
are illustrated in Fig.3 by surfaces ${\cal S}_1,{\cal S}_2$ and
${\cal S}_3$, respectively.  If, for example, one uses the surface
${\cal S}_2$ as the cutoff surface, one would conclude that all
observers will see the same vacuum with 100\% probability.  With a
suitable choice of the surface, one can get any result for the
relative probability of the two minima.

\begin{figure}[t]
\psfig{file=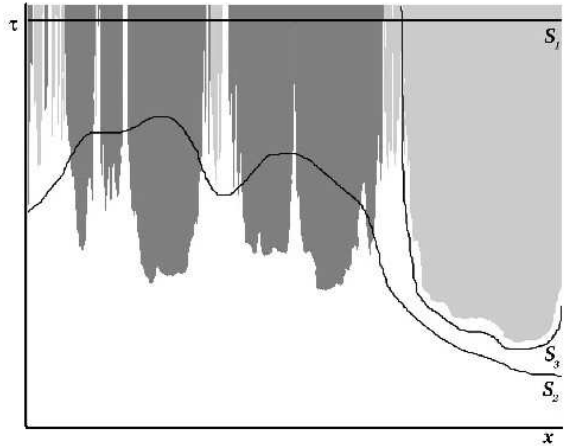,width=3in,angle=0}
\caption{${\cal S}_1$ is a surface of a constant proper time.
It crosses many thermalized regions of different types.  ${\cal S}_2$
is a spacelike surface which crosses regions of only one type.  ${\cal
S}_3$ is a spacelike surface which does not cross any thermalized
regions.}
\label{fig.3}
\end{figure}

We have also performed simulations for a two-field model with a
potential
\beq
V(\vp,\chi)=V_0\cos^2(\kappa\vp)[1+\lambda(1+\cos\beta\chi)
\sin^4\kappa\vp],
\label{Vphichi}
\eeq
where $V_0,\kappa,\beta$ and $\lambda$ are constants.  We start the
simulation in a horizon-size region with $\vp=\chi=0$ at the initial
moment $t=0$.  A spacetime slice through one of the simulations is
given in Fig.4, where different values of $\chi$ are shown with
different shades of grey in the inflating regions, while the
thermalized regions are left white.  The field $\chi$ is homogeneous
at the bottom of the figure ($t=0$), but gets randomized by quantum
fluctuations as time goes on.  Different regions of space thermalize
with different values of $\chi$, and $\chi$ takes all of its values on
each thermalization surface.  It is clear that the simple explanation
of gauge-dependence that I gave for the one-field model, 
that constant-$t$ surfaces can be chosen so that
they cross one type of thermalized regions and avoid the other, does
not apply in this case.  For the two-field model, the situation is
more subtle.  

\begin{figure}[t]
\psfig{file=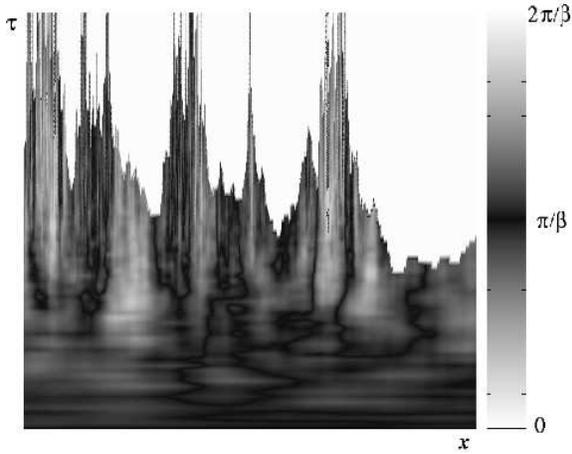,width=3in,angle=0}
\caption{Spacetime structure in a one-dimensional simulation
for the two-field model showing the evolution of a comoving region of
initial size $l=H^{-1}$ with initially homogeneous $\phi=0$ and
$\chi=\pi/\beta$.  The values of the field $\chi$ are shown
throughout the inflating region using different shades of grey.  The
shading code is indicated in the bar on the right of the figure.  The
thermalized regions are left white.}
\label{fig.4}
\end{figure}

Note that the thermalized regions in Fig.4 have a
fractal-like pattern, with larger numbers of increasingly narrow
``spikes'' appearing at later times.  These spikes represent
newly-formed thermalized regions, and although they are rather small,
their number grows exponentially with time, and
they actually dominate the thermalized volume.  A change of the time
variable $t$ results in a deformation of the cutoff surface,
accompanied by a (substantial) change in the population of the
newly-formed regions that are being included.  This is the origin of
the gauge dependence of the cutoff procedure.

\section{The proposal}

The resolution of the gauge dependence problem that I proposed in
Ref.\cite{AV98} is to calculate the probability distribution for
$\chi$ within a
single, connected thermalized domain.  Each thermalized domain can be
infinitely extended and 
contains an infinite number of galaxies, but it is
sufficient to use a large finite part of the domain.  If the field
$\chi$ varies in a finite range, it will run through all of its values
many times in a sufficiently large volume.  We expect, therefore, 
that the distribution $F(\chi)$ will converge rapidly as the
volume is increased.  It does not matter which thermalized domain we
choose to calculate probabilities: all domains are statistically
equivalent, due to the stochastic nature of quantum fluctuations in
eternal inflation.  This is a very simple prescription, and I am a bit
embarrassed that I did not think of it earlier, having thought about
this problem for a number of years.

With this prescription, the volume distribution $F(\chi)$ can be
calculated directly from numerical simulations, and we have done that
in \cite{VVW} for the two-field model (\ref{Vphichi}).  In some cases an
analytic calculation is also possible.  Suppose, for example, that the
potential $V(\vp,\chi)$ is essentially independent of $\chi$ for
$|\vp|<\vp_0$, where $\vp_0$ is in the deterministic slow-roll range,
where quantum fluctuations of $\phi$ and $\chi$ can be neglected
compared to the classical drift.   (The corresponding conditions on the
parameters of the potential (\ref{Vphichi}) are specified in
\cite{AV98,VVW}). Then, the evolution of $\vp$ at $\vp>\vp_0$ is
monotonic, and a natural choice of the time variable in this range is
$t=\vp$.  The probability distribution for $\chi$ on the
constant-''time'' surface $\vp=\vp_0$ is
\beq
F_0(\chi)=F(\vp_0,\chi)=const,
\label{flat}
\eeq
since all values of $\chi$ are equally probable at $\vp<\vp_0$.  We
are interested in the probability distribution on the thermalization
surface, $F(\chi)=F(\vp_*,\chi)$, where $\vp_*$ is the value of $\vp$
at thermalization.  This is given by \cite{AV98}
\beq
F(\chi)\propto F_0(\chi_0)\exp[3N(\chi_0)]det\left|{{\p\chi_0}\over{\p\chi}}
\right|.
\label{anal}
\eeq
Here, $\chi_0$ is the value of $\chi$ at $\vp=\vp_0$ that classically
evolves into $\chi$ at $\vp_*$, $N(\chi_0)$ is the number of e-foldings along
this classical path, $\exp(3N)$ is the corresponding enhancement of
the volume, and the last factor is the Jacobian transforming from
$\chi_0$ to $\chi$.  In many interesting cases, $\chi$ does not change
much during the slow roll.  Then,
\beq
F(\chi)\propto \exp[3N(\chi)].
\eeq

In a more general case, when the diffusion of $\chi$ is not negligible
at $\vp>\vp_0$, the distribution $F(\chi)$ can be found by solving the
Fokker-Planck equation with $t=\vp$ in the range $\vp_0<\vp<\vp_*$ and
with the initial condition (\ref{flat}).  The corresponding form of
the equation was derived in \cite{VVW},
\beq
{\p F\over{\p\vp}}=-\p_\chi^2\left({H^3 F\over{2\pi {H'}_\vp}}\right)
-\p_\chi\left({{H'}_\chi\over{{H'}_\vp}}F\right) -{12\pi
H\over{{H'}_\vp}}F. 
\eeq
We have solved this equation for the two-field model (\ref{Vphichi})
with the same parameters that we used in numerical simulations and
compared the resulting probability distribution $F(\chi)$ with the
distribution obtained directly from the simulations.  We found very
good agreement between the two (see Ref.\cite{VVW} for details).

\section{Density fluctuations}

As a specific application of the proposed approach, let us consider
the spectrum of density perturbations in the standard model of
inflation with a single field $\vp$.  The perturbations are determined
by quantum fluctuations $\delta\vp$; they are introduced on each
comoving scale at the time when that scale crosses the horizon and
have a gauge-invariant amplitude \cite{Mukhanov}
\beq
\delta\rho/\rho=8\pi H\delta\vp/H',
\eeq
where $H'=dH/d\vp$.  With an rms fluctuation
$(\delta\vp)_{rms}=H/2\pi$, this gives
\beq
(\delta\rho/\rho)_{rms}=4H^2/|H'|.
\label{drr}
\eeq
Fluctuations of $\vp$ on different length scales
are statistically independent and can be treated separately.  We can
therefore concentrate on a single scale corresponding to some value
$\vp=\vp_0$, disregarding all of the rest.  

On the equal-''time'' surface $\vp=\vp_0$, the fluctuations
$\delta\vp$ can be regarded as random Gaussian variables with a
distribution
\beq
F_0(\delta\vp)\propto\exp\left[-{2\pi^2\over{H_0^2}}(\delta\vp)^2\right],
\label{F1}
\eeq
where $H_0=H(\vp_0)$.  We are interested in the distribution
$F(\delta\vp)$ on the thermalization surface $\vp=\vp_*$.  This will
be different from $F_0$ if there is some correlation between
$\delta\vp$ and the amount of inflationary expansion in the period
between $\vp_0$ and $\vp_*$.  In fact, there is such a correlation.
If $\vp$ fluctuates in the direction opposite to the classical roll, 
then inflation is prolonged and the expansion factor is increased.
Otherwise, it is decreased, and we can write
\beq
F(\delta\vp)\propto F_0(\delta\vp)\exp(3H_0\delta t),
\label{F2}
\eeq
where 
\beq
\delta t=-(4\pi/{H'}_0)\delta\vp
\label{F3}
\eeq
is the time delay of the slow roll due to the fluctuation $\delta\vp$.  

Combining Eqs.(\ref{F1})-(\ref{F3}), we obtain \cite{AV98}
\beq
F(\delta\vp)\propto\exp\left[-{2\pi^2\over{H_0^2}}(\delta\vp-{\bar
{\delta\vp}})^2 \right],
\eeq
which describes Gaussian fluctuations with a nonzero mean value,
\beq
{\bar{\delta\vp}}=3H_0^3/\pi {H'}_0.
\eeq 
This is different from the standard approach \cite{Mukhanov} which
disregards the volume enhancement factor and uses the distribution
(\ref{F1}).  The effect, however, is hopelessly small.  Indeed,
\beq
{{\bar{\delta\vp}}\over{(\delta\vp)_{rms}}}={6H_0^2\over{{H'}_0}}\sim 
\left({\delta\rho\over{\rho}}\right)_{rms}\sim 10^{-5}.
\eeq
We thus see that the standard results remain essentially unchanged.

\section{Conclusions}

Eternally inflating universes can contain thermalized regions with
different values of the cosmological parameters, which we have denoted
generically by $\chi$.  We cannot then predict $\chi$ with certainty
and can only find the probability distribution ${\cal P}(\chi)$.  
Until recently, it was thought that calculation of ${\cal P}$
inevitably involves comparing infinite volumes, and therefore leads to
ambiguities.  My proposal is to calculate ${\cal P}$ in a single
thermalized domain.  The choice of the domain is unimportant, since
all thermalized domains are statistically equivalent.  This apprach
gives unambiguous results.  When applied to the spectrum of density
fluctuations, it recovers the standard results with a small correction
$O(10^{-5})$.  

It should be noted that this approach cannot be applied to models
where $\chi$ is a discrete
variable which takes different values in different thermalized
regions, but is homogeneous within each region.  [An example of such a
model is the one-field model (\ref{Vphi}).]  One can take this as 
indicating that no probability distribution for a discrete variable
can be meaningfully defined in an eternally inflating universe.
Alternatively, one could try to introduce some other cutoff
prescription to be applied specifically in the case of a discrete
variable.  Some possibilities have been discussed in
\cite{AV95,VW96}.  This issue requires further investigation.

\end{document}